\documentclass[a4paper,prd,nofootinbib,preprint,showpacs,showkeys]{revtex4}

\usepackage[english]{babel}
\usepackage{amsmath,amssymb}
\usepackage[dvips]{graphicx}

\newcommand{\pnu}[1]{\overset{\smash{\scriptscriptstyle (-)}}{\nu}_{\hskip-3pt #1}}
\newcommand{\epS}{\epsilon_S}
\newcommand{\epP}{\epsilon_P}
\newcommand{\si}{S_{13}}

\begin{document}

\preprint{IFIC/02--06}
\preprint{TUM--hep 451/02}
\preprint{MPI--Pht/2002--04}

\title{\vspace*{1 cm} Confusing non-standard neutrino interactions
  with oscillations at a neutrino factory}

\author{P.~Huber} \email{Patrick.Huber@ph.tum.de}

\affiliation{Theoretische Physik, Physik Department, 
  Technische Universit{\"a}t M{\"u}nchen,\\
  James--Franck--Strasse, D--85748 Garching, Germany\\
  Max-Planck-Institut f{\"u}r Physik, Postfach 401212, 
  D--80805 M{\"u}nchen, Germany}

\author{T.~Schwetz} \email{schwetz@ific.uv.es} 

\author{J.W.F.~Valle} \email{valle@ific.uv.es}

\affiliation{Instituto de F\'{\i}sica Corpuscular -- C.S.I.C., 
  Universitat de Val{\`e}ncia \\
  Edificio Institutos, Aptdo.\ 22085, E--46071 Val{\`e}ncia,
  Spain\vspace*{1cm}}

\begin{abstract}
  Most neutrino mass theories contain non-standard interactions (NSI)
  of neutrinos which can be either non-universal (NU) or
  flavor-changing (FC).  We study the impact of such interactions on
  the determination of neutrino mixing parameters at a neutrino
  factory using the so-called ``golden channels''
  $\pnu{e}\to\pnu{\mu}$ for the measurement of $\theta_{13}$. We show
  that a certain combination of FC interactions in neutrino source and
  earth matter can give exactly the same signal as oscillations
  arising due to $\theta_{13}$.  This implies that information about
  $\theta_{13}$ can only be obtained if bounds on NSI are available.
  Taking into account the existing bounds on FC interactions, this
  leads to a drastic loss in sensitivity in $\theta_{13}$, at least
  two orders of magnitude. A near detector at a neutrino factory
  offers the possibility to obtain stringent bounds on some NSI
  parameters. Such near site detector constitutes an essential
  ingredient of a neutrino factory and a necessary step towards the
  determination of $\theta_{13}$ and subsequent study of leptonic CP
  violation.
\end{abstract}



\maketitle

\section{Introduction}

From the long-standing solar \cite{solar} and atmospheric \cite{atmos}
neutrino anomalies we now have compelling evidence that an extension of the
Standard Model (SM) of particle physics is necessary in the lepton sector. The
simplest and most generic explanation of these experiments is provided by
neutrino oscillations induced by neutrino masses and mixing.  As is well known
and accepted, the indication of the LSND experiment \cite{LSND} for
oscillations at a large mass-squared difference can not be reconciled with
solar and atmospheric data within a 3-neutrino framework.  For recent
four-neutrino~\cite{Peltoniemi:1992ss} analyses see Ref.~\cite{Maltoni:2001bc}
and references therein.  For this reason we choose not to consider the LSND
data and focus therefore on the simplest 3-neutrino scheme with the two
mass-squared differences $\Delta m^2_{\mathrm{sol}} \lesssim 10^{-4}$ eV$^2$
and $\Delta m^2_{\mathrm{atm}} \approx 3\times 10^{-3}$
eV$^2$~\cite{Gonzalez-Garcia:2001sq}. The corresponding lepton mixing matrix
is parameterized by the three angles $\theta_{12},\,\theta_{23},\,\theta_{13}$
and one complex CP-violating phase $\delta$ (see later for exact definitions)
relevant in lepton-number-conserving neutrino oscillations
\cite{theory,theory1}. It is known from atmospheric neutrino data that
$\theta_{23}$ has to be nearly maximal.  On the other hand, data from present
solar neutrino experiments favor a large value for the angle $\theta_{12}$
\cite{Gonzalez-Garcia:2000aj,solaranalysis}.  An improved determination is
expected from solar neutrino data and/or the results of the KamLAND experiment
\cite{kamland}. The value of the third angle $\theta_{13}$ is not known at
present, there is only the bound
\begin{equation}\label{choozbound}
\sin^2 2\theta_{13} \lesssim 0.1 \quad\mbox{at}\quad 90\%\:\mbox{CL}
\end{equation}
implied by combining the results of the reactor
experiments~\cite{reactor} CHOOZ and Palo Verde with 
atmospheric data. Together with the requirement of solar neutrino 
oscillations this bound implies that $\sin^2 \theta_{13}$ has to 
be small. The value of the phase $\delta$ is completely
unknown.

Currently a new generation of long-baseline neutrino oscillation
experiments using a neutrino beam originating from the decay of muons
in a storage ring is being discussed \cite{nuFacPapers}.  These
so-called {\sl neutrino factories} are considered as the ideal tool to
enhance our knowledge about neutrino mixing parameters. Besides the
possibility to explore CP violation in the lepton sector an important
aim of a neutrino factory will be a precise determination of
$\theta_{13}$; it is claimed that a measurement of $\theta_{13}$ down
to values of a few$\times 10^{-4}$ will be
possible~\cite{Freund:2001ui}.

In a large class of models beyond the Standard Model non-standard
interactions (NSI) of neutrinos with matter arise. 
The simplest NSI do not require new interactions beyond those
mediated by the Standard Model electroweak gauge bosons: it is simply
nature of the leptonic charged and neutral current interactions which
is non-standard due to the complexity of neutrino mixing
\cite{theory}. On the other hand NSI can also be mediated by the
exchange of new particles with mass at the weak scale such as in some
super-symmetric models with R-parity violating
\cite{Ross:1985yg,Hall:1984id} interactions.
Such non-standard flavor-violating physics can arise even in the
absence of neutrino mass \cite{NSImodels2,NSImodels3} and can
lead to non-universal (NU) or to flavor-changing (FC) neutrino
interactions.

Non-standard interactions of neutrinos affect their propagation in
matter and the magnitude of the effect depends on the interplay
between conventional mass-induced neutrino oscillation features in
matter \cite{MSW} and those genuinely engendered by the NSI, which do
not require neutrino mass \cite{Valle:1987gv}.
Correspondingly their implications have been explored in a variety of
contexts involving solar neutrinos
\cite{MSW,Valle:1987gv,NSIrecent,Bergmann:2000gp,Guzzo:2001mi},
atmospheric neutrinos \cite{Guzzo:2001mi,NSIatm,Val,BergmannAtm},
other astrophysical sources \cite{NSIastro,NSIastro1} and the LSND
experiment \cite{BergmannLSND,Bergmann:2000gn}. The impact of non-standard
interactions of neutrinos has also been considered from the point of
view of future experiments involving solar neutrinos
\cite{Berezhiani:2001rt} as well as the upcoming neutrino factories
\cite{Huber:2001zw,Huber:2001de}.  Various aspects of NSI for a
neutrino factory experiment have been considered in
Refs.~\cite{Huber:2001zw,Huber:2001de,Gago:2001xg,Gonzalez-Garcia:2001mp,Datta:2000ci,johnson,ota,Bueno:2001jy}.

In this paper we will consider from a phenomenological point of view
the impact of NSI on the determination of neutrino mixing parameters
at a neutrino factory.  In particular we will focus on the
$\pnu{e}\to\pnu{\mu}$ channels, which are supposed to be the ``golden
channels'' for the measurement of $\theta_{13}$.  We extend our previous
work \cite{Huber:2001de} by taking simultaneously into account
neutrino oscillations and the effect of NSI in neutrino source,
propagation and detection~\cite{ota}. We will show that a certain
combination of FC interactions in source and propagation can give
exactly the same signal as oscillations arising due to $\theta_{13}$.
This implies that information about $\theta_{13}$ can only be obtained
if bounds on NSI are available.  In view of the existing bounds on
FC interactions, this leads to a drastic loss in sensitivity in $\theta_{13}$,
at least two orders of magnitude.

All our considerations also apply to the determination of
$\theta_{13}$ in long baseline experiments using upgraded conventional
beams~\cite{superbeams,ota}. However, due to the different production
processes involved, the NSI parameters relevant in the source of a
conventional beam experiment differ from the ones at a neutrino
factory. The same methods discussed here can be adapted to
cover also that case. A detailed numerical consideration of conventional
beam experiments goes beyond the scope of this work.

The outline of the paper is as follows.  In section \ref{sec:models}
we briefly sketch the theoretical motivation for NSI in the context of
gauge theories of neutrino mass. In section \ref{sec:framework} we
discuss examples of low energy four-fermion Hamiltonians, which lead
to NSI in neutrino source, propagation and detection.  In section
\ref{sec:bounds} we review some bounds on NSI parameters obtained in
the literature. In section \ref{sec:rate} we present the framework of
our numerical calculations and discuss the appearance rate in the
presence of NSI and oscillations. In section \ref{sec:confusion} we
derive analytical expressions for this rate and formulate the {\sl
  oscillation--NSI confusion theorem}.  In section
\ref{sec:simulation} we describe the simulation of a neutrino factory
and our statistical method to investigate the possibilities of such an
experiment to determine NSI and oscillation parameters.  In section
\ref{sec:sensitivity} we define sensitivity limits for
$\sin^22\theta_{13}$ and show our numerical results for the three
baselines 700 km, 3000 km and 7000 km as a function of bounds on the
relevant NSI parameters.  We also discuss the sensitivity limits if
information from two different baselines is combined.  Finally we
conclude in section \ref{sec:conclusions}.

\section{Theoretical motivation}
\label{sec:models}

More often than not, models of neutrino mass are accompanied by NSI,
leading generically to both oscillations and neutrino NSI in matter.
The simplest are those NSI which arise from neutrino-mixing.  The most
straightforward example of this case is when neutrino masses follow
from the admixture of isosinglet neutral heavy leptons as, for
example, in seesaw schemes~\cite{GRS}. These contain $SU(2) \otimes U(1)$
singlets with a gauge invariant Majorana mass term of the type
${M_R}_{ij} \nu^c_i \nu^c_j$ which breaks total lepton number symmetry,
perhaps at a large $SO(10)$ or left-right breaking scale.  The masses
of the light neutrinos are obtained by diagonalizing the mass matrix
\begin{equation}
    \label{eq:SS}
    \begin{bmatrix}
        M_L & D \\
        D^T & M_R
    \end{bmatrix}
\end{equation}
in the basis $\nu,\nu^c$, where $D$ is the standard $SU(2) \otimes U(1)$
breaking Dirac mass term, and $M_R = M_R^T$ is the large isosinglet
Majorana mass and the $M_L \nu\nu$ term is an isotriplet~\cite{theory}.
In left-right models the latter is generally suppressed as $M_L \propto
1/M_R$.

The structure of the associated effective weak $V-A$ currents is rather
complex~\cite{theory}. The first point to notice is that the
heavy isosinglets will mix with the ordinary isodoublet neutrinos in
the charged current weak interaction. As a result, the mixing matrix
describing the charged leptonic weak interaction is a rectangular
matrix $K$~\cite{theory} which may be decomposed as
\begin{equation}
    \label{eq:CC}
    K = (K_L, K_H)
\end{equation}
where $K_L$ and $K_H$ are $3 \times 3$ matrices.  The corresponding neutral
weak interactions are described by a non-trivial matrix~\cite{theory}
\begin{equation}
    \label{eq:NC}
    P = K^\dagger K \, .
\end{equation}

In such models non-standard interactions of neutrinos with matter
arise from the non-trivial structures of the charged and neutral weak
currents.  Note, however, that the smallness of neutrino mass,
which follows due to the seesaw mechanism $M_{\nu \: \mathrm{eff}} = M_L - D
M_R^{-1} D^T$ and the condition
\begin{equation}
    M_L \ll M_R \,,
\end{equation}
implies that the magnitude of neutrino NSI is expected to be
negligible. However this need not be so in general.  For example,
since the number $m$ of $SU(2) \otimes U(1)$ singlets is arbitrary, one
may consider models with Majorana neutrinos based on {\sl any value}
of $m$. One can therefore extend the lepton sector of the $SU(2)
\otimes U(1)$ theory by adding a set of {\sl two} 2-component isosinglet
neutral fermions, denoted ${\nu^c}_i$ and $S_i$, in each generation
\cite{NSImodels2}. In such $m = 6$ models one can consider the $9
\times 9$ mass matrix~\cite{Gonzalez-Garcia:1989rw}
\begin{equation}
    \label{eq:MATmu}
    \begin{bmatrix}
        0 & D & 0 \\
        D^T & 0 & M \\
        0 & M^T & \mu
    \end{bmatrix}
\end{equation}
(in the basis $\nu, \nu^c, S$). The Majorana masses for the neutrinos are
determined from
\begin{equation}
    \label{eq:33}
    M_L = D M^{-1} \mu {M^T}^{-1} D^T \, .
\end{equation}
In the limit $\mu \to 0$ the exact lepton number symmetry is recovered
and neutrinos become massless~\cite{NSImodels2}.
This provides an elegant way to generate neutrino masses without a
super-heavy scale and automatically allows one to enhance the 
magnitude of neutrino NSI strengths by avoiding the constraints which
arise from the smallness of neutrino masses presently indicated by the
oscillation interpretation of solar and atmospheric neutrino data.

The propagation of the light neutrinos is effectively described by a
truncated mixing matrix $K_L$ which is not unitary. This may lead to
oscillation effects in matter, even if neutrinos were
massless~\cite{Valle:1987gv}. They maybe be resonant and therefore
important in supernovae matter \cite{Valle:1987gv,NSIastro}.
The strength of NSI is hence unrestricted by the magnitude of neutrino
masses, only by universality limits, and may be large, at the few per cent
level.  The phenomenological implications of these models have been widely
investigated~\cite{Bernabeu:1987gr,Gonzalez-Garcia:1990fb,Rius:1990gk,Valle:1991pk,Gonzalez-Garcia:1992be}.

An alternative way to induce neutrino NSI is in the context of the
most general low-energy super-symmetry model, without R-parity
conservation~\cite{Ross:1985yg}. In addition to bilinear violation
\cite{Hall:1984id,Hirsch:2000ef} one may also have trilinear $L$
violating couplings in the super-potential
\begin{gather}
    \label{eq:lq}
    \lambda_{ijk} L_i L_j E^c_k \, \\
    \lambda'_{ijk} L_i Q_j D^c_k
\end{gather}
where $L, Q, E^c$ and $D^c$ are (chiral) super-fields which contain the usual
lepton and quark $SU(2)$ doublets and singlets, respectively, and $i,j,k$ are
generation indices.
The couplings in Eq.~(\ref{eq:lq}) give rise at low energy to the following
four-fermion effective Lagrangian for neutrinos interactions with $d$-quark
including
\begin{equation}
    \label{eq:effec}
   \mathcal{L}_\mathrm{eff}  =  - 2\sqrt{2} G_F \sum_{\alpha,\beta}
    \xi_{\alpha\beta} \: \bar{\nu}_{L\alpha} \gamma^{\mu} \nu_{L\beta} \:
    \bar{d}_{R}\gamma^{\mu}{d}_{R}\:\:\:\alpha,\beta = e,\mu, \tau \, ,
\end{equation}
where the parameters $\xi_{\alpha\beta}$ represent the strength of the
effective interactions normalized to the Fermi constant $G_F$.
One can identify explicitly, for example, the following {\sl non-standard}
flavor-conserving NSI couplings
\begin{align}
    \xi_{\mu\mu} &= \sum_j \frac{|\lambda'_{2j1}|^2}
    {4 \sqrt{2} G_F m^2_{\tilde{q}_{j L}}} \, , \\
    \xi_{\tau\tau}& = \sum_j \frac{|\lambda'_{3j1}|^2}
    {4 \sqrt{2} G_F m^2_{\tilde{q}_{j L}} }\, ,
\end{align}
and the FC coupling
\begin{equation}
    \xi_{\mu\tau} =  \sum_j \frac{ \lambda^\prime_{3j1} \lambda^\prime_{2j1} }
    {4\sqrt{2}G_F m^2_{\tilde{q}_{jL}} }
\end{equation}
where $m_{\tilde{q}_{j L}}$ are the masses of the exchanged squarks and $j =
1,2,3$ denotes $\tilde{d}_L, \tilde{s}_L, \tilde{b}_L$, respectively.
The existence of effective neutral current interactions contributing
to the neutrino scattering off $d$-quarks in matter, provides new
flavor-conserving as well as flavor-changing terms for the matter
potentials of neutrinos.  Such NSI are directly relevant for solar and
atmospheric neutrino propagation \cite{Guzzo:2001mi}.  

Clearly, such neutrino NSI are accompanied by non-zero neutrino
masses. In fact one has a hybrid model for neutrino masses in which
the atmospheric scale arises at the tree level while the solar
neutrino scale is induced by loops which involve directly the NSI
coefficients in (\ref{eq:lq}). This way one obtains the co-existence
of oscillations as well as NSI of neutrinos. The relative
importance of NSI and oscillation features is parameter-dependent.

An alternative variant of the above scheme is provided by some
radiative models of neutrino masses such as the one discussed
in~\cite{Fukugita:1988qe}. In all such models NSI may arise from
scalar interactions.

Finally, we mention that unification provides an alternative and
elegant way to induce neutrino NSI. For example unified super-symmetric
models lead to NSI as a result of super-symmetric scalar lepton
non-diagonal vertices induced by renormalization group
evolution~\cite{NSImodels3,Barbieri:1995tw}.  In the special case of
$SU(5)$ the NSI may exist without neutrino mass. In $SO(10)$ neutrino
masses co-exist with neutrino NSI.

In what follows we shall investigate the interplay of NSI-induced and
neutrino-mass-induced (oscillation-induced) conversion of neutrinos at a
neutrino factory and on how it can vitiate the, otherwise very
precise, determination of neutrino oscillation parameters.

\section{Effective four-fermion Hamiltonians describing NSI}
\label{sec:framework}

In this section we consider in some detail the simplest examples of
effective low energy Hamiltonians, which lead to NSI in the source
($S$), propagation ($P$) and detection ($D$) of neutrinos in a
neutrino factory experiment. 

In such an experiment neutrinos are produced by the decay of the
stored muons $\mu^+ \to e^+ + \nu_e + \bar\nu_\mu$ and the charge
conjugated process. In the SM these processes are described by the
effective four-fermion Hamiltonian
\begin{equation}\label{SMS}
\mathcal{H}_{\mathrm{SM}}^S= \frac{G_F}{\sqrt{2}}
\left[ \bar\nu_\mu \,(1-\gamma_5)\gamma_\lambda \, \mu \right]
\left[\bar e \,(1-\gamma_5)\gamma^\lambda \, \nu_e \right] +\mathrm{h.c.}
\end{equation}
In addition to this SM term, resulting from the exchange of the
W-boson, we consider now new processes $\mu^+ \to e^+ + \nu_\alpha +
\bar\nu_\mu$ with any flavor $\alpha=e,\mu,\tau$ for the neutrino
related to the positron.\footnote{In this work we will consider the
  $e\to\mu$ appearance channel and hence, we are interested only in
  the neutrino produced together with the positron (or the electron,
  for the charge conjugated processes).  More generally, also
  processes $\mu^+ \to e^+ + \nu_\alpha + \bar\nu_\beta$ with an
  arbitrary flavor combination $(\alpha,\beta)=(e,\mu,\tau)$ may
  occur. In this case the final state in the source can be different 
  from the one in the SM. Such additional processes must be
  added incoherently to obtain the transition rate defined later in
  Eq.~(\ref{rate}). For simplicity we will not consider this
  possibility further.}
We parametrize the corresponding NSI effective Hamiltonian by the
coefficients $\epsilon^S_{e\alpha}$:
\begin{equation}\label{NSIS}
\mathcal{H}_{\mathrm{NSI}}^S = \frac{G_F}{\sqrt{2}}
\left[ \bar\nu_\mu \,(1-\gamma_5)\gamma_\lambda \, \mu \right]
\sum_\alpha \epsilon^S_{e\alpha}
\left[ \bar e \,(1-\gamma_5)\gamma^\lambda \, \nu_\alpha \right]
+\mathrm{h.c.}
\end{equation}

Note that the value of the SM Fermi constant $G_F$ is determined
experimentally from the decay-width of the muon~\cite{pdg}, without
measuring the flavor of the neutrinos. Therefore, we have the relation
\begin{equation}\label{GFexp}
G_F^{\mathrm{exp}}  =
G_F \left( |1+\epsilon^S_{ee}|^2 +
\sum_{\alpha=\mu,\tau} |\epsilon^S_{e\alpha}|^2 \right)^{1/2} \,.
\end{equation}
The term $\epsilon^S_{ee}$ leads to exactly the same final state as
the SM process and must be added coherently, whereas
$\epsilon^S_{e\mu}$ and $\epsilon^S_{e\tau}$ lead to different final
states and contribute incoherently to the decay. From
Eq.~(\ref{GFexp}) one learns that the high precision measurement of
$G_F^{\mathrm{exp}}$ on its own (within an accuracy of
$9\times10^{-6}$~\cite{pdg}) does not constrain any of the parameters
in the Hamiltonian $G_F$ and $\epsilon^S_{e\alpha}$
directly~\cite{Langacker:1988cm}; only the {\sl combination} shown in
Eq.~(\ref{GFexp}) is constrained within the accuracy of the
experimental measurement.

The standard muon detectors under discussion for a neutrino factory
experiment make use of charged current processes like
$\nu_\mu+d\to\mu^- + u$. The relevant effective Hamiltonian in the SM
is given by
\begin{equation}\label{SMD}
\mathcal{H}_{\mathrm{SM}}^D= \frac{G_F}{\sqrt{2}}
\left[ \bar d \,(1-\gamma_5)\gamma_\lambda \, u \right]
\left[\bar \nu_\mu \,(1-\gamma_5)\gamma^\lambda \, \mu \right] +\mathrm{h.c.}
\end{equation}
Here $d\:(u)$ symbolizes any down-(up-)type quark.  Similar to
Eq.~(\ref{NSIS}) we consider the following NSI four-fermion
Hamiltonian:
\begin{equation}\label{NSID}
\mathcal{H}_{\mathrm{NSI}}^D = \frac{G_F}{\sqrt{2}}
\left[ \bar d \,(1-\gamma_5)\gamma_\lambda \, u \right]
\sum_\alpha \epsilon_{\alpha\mu}^D
\left[\bar \nu_\alpha \,(1-\gamma_5)\gamma^\lambda \, \mu \right] 
+\mathrm{h.c.}
\end{equation}
The coefficients $\epsilon_{\alpha\mu}^D$ describe NU ($\alpha=\mu$) or FC ($\alpha=e,\tau$)
NSI in the detector, {\sl e.g.}~a non-zero $\epsilon_{\tau\mu}^D$ leads to the
process $\nu_\tau + d \to \mu^- + u$.

In a long-baseline neutrino experiment a significant part of the
neutrino path will cross the earth and hence neutrino NSI with earth
matter must be taken into account. Let us consider the effective
Hamiltonian describing the SM neutral current processes of neutrinos
with a fermion $f$ due to the exchange of the Z-boson $\nu_\alpha + f
\to \nu_\alpha + f$:

\begin{equation}\label{SMP}
\mathcal{H}_{\mathrm{SM}}^P = \frac{G_F}{\sqrt{2}}
\left[ \bar f \,(g_V^f - g_A^f\gamma_5)\gamma_\lambda \, f \right]
\sum_\alpha
\left[\bar \nu_\alpha \,(1-\gamma_5)\gamma^\lambda \, \nu_\alpha \right] \,,
\end{equation}
where $g_V^f$ and $g_A^f$ are the SM vector and axial couplings of the fermion
$f$, see {\sl e.g.}~Ref.~\cite{pdg} Sec.~10.  In the SM this interaction is
the same for all flavors and hence, has no effect on the propagation of the
neutrino state -- in contrast to the charged current interaction of $\pnu{e}$
with electrons. However, if NSI are present, we must also take into account
processes $\nu_\alpha + f \to \nu_\beta + f$ with arbitrary flavor
combinations ($\alpha\beta$) described by the Hamiltonian

\begin{equation}\label{NSIP}
\mathcal{H}_{\mathrm{NSI}}^P = \frac{G_F}{\sqrt{2}}
\left[ \bar f \,(g_V^f - g_A^f\gamma_5)\gamma_\lambda \, f \right]
\sum_{\alpha\beta} \epsilon^f_{\alpha\beta}
\left[\bar \nu_\alpha \,(1-\gamma_5)\gamma^\lambda \, \nu_\beta \right] \,.
\end{equation}

In Eqs.~(\ref{NSIS}), (\ref{NSID}), (\ref{NSIP}) we have assumed for
simplicity, that the NSI have the same $V-A$ Lorentz structure as the
SM interactions.  This needs not to be the case in the most general
extension of the SM involving, say, left-right symmetry, where many
new NSI parameters can appear (see {\sl e.g.}
Ref.~\cite{johnson,ota}).  However, the effects of NSI with $V+A$
Lorentz structure are strongly suppressed since the left-right
breaking scale should be rather high in order to account for the
smallness of the neutrino masses indicated by solar and atmospheric
experiments.  Moreover, one expects that only certain combinations of
parameters will be relevant for the experimental configuration we are
considering here, and for any given theory our results can be mapped
to the corresponding combination of parameters.

Although different processes are relevant for source, propagation and
detection, in a given model relations between the coefficients
$\epsilon^X_{\alpha\beta}$ may exist. However, such
relations highly depend on the underlying model.  In order to be
model-independent we will treat all $\epsilon^X_{\alpha\beta}$ as
independent parameters.

\section{Bounds on NSI parameters}
\label{sec:bounds}

In this section we review existing bounds on neutrino NSI obtained in the
literature. The most direct upper bounds on the strength of NSI interactions
arise from negative searches for neutrino
oscillations~\cite{johnson,Grossman:1995wx,Gninenko:2001id}. For example the
bounds on the transition probabilities $P_{\nu_\mu\to\nu_\tau} \leq 3.4\times
10^{-4}$ and $P_{\nu_e\to\nu_\tau} \leq 2.6\times 10^{-2}$ obtained by CHORUS
\cite{chorus} yield the bounds
\begin{equation}\label{boundChorus}
|\epsilon_{\mu\tau}^{\mathrm{CHORUS}}| \lesssim 1.8 \times 10^{-2} \,,\quad
|\epsilon_{e\tau}^{\mathrm{CHORUS}}| \lesssim 0.16\,.
\end{equation}
With the superscript we indicate that the constrained quantity
actually is a certain combination of NSI coefficients relevant in the
neutrino source and detection processes for a given experiment (see
later Eq.~(\ref{rateND})), which in general is different from the NSI
coefficients relevant for neutrino factory experiments.

Recently, in Ref.~\cite{Val} the strong evidence for oscillations of
atmospheric neutrinos has been used to set upper bounds on NSI of neutrinos
with the down quarks in earth matter:
\begin{equation}\label{boundAtm}
|\epsilon_{\mu\tau}^d| \lesssim 3 \times 10^{-2} \,,\quad
|\epsilon_{\tau\tau}^d| \lesssim 6 \times 10^{-2}\,.
\end{equation}
Similar bounds on the magnitude of neutrino NSI with electrons and up-type
quarks may be derived~\cite{Guzzo:2001mi}.

Besides these direct bounds on neutrino NSI there is a lot of data
constraining non-standard effects in {\sl charged lepton} processes.
However, it is very non-trivial in general to use these data to derive
model-independent bounds on neutrino NSI coefficients. For a recent
discussion see, for example
Refs.~\cite{BergmannAtm,BergmannLSND,Gonzalez-Garcia:2001mp}.  To
obtain such bounds for neutrino interactions one must convert from the
bounds on charged lepton processes making some assumption about
$SU(2)_L$ symmetry.  In Ref.~\cite{Bergmann:2000gp} the corresponding
bounds for the charged leptons are multiplied by a factor of $\approx
6.8$, in order to take into account $SU(2)_L$ breaking effects.  In
this way the following bounds on neutrino NSI are derived from pure
leptonic processes~\cite{Bergmann:2000gp}:
\begin{equation}\label{boundsLep} 
|\epsilon^\ell_{e\mu}| \lesssim 7\times 10^{-6}\,,\quad
|\epsilon^\ell_{e\tau}| \lesssim 3\times 10^{-2}\,.
\end{equation}
As the neutrino production in a neutrino factory is also a pure
leptonic process we take the bounds (\ref{boundsLep}) as order of
magnitude estimates of the NSI at the neutrino source. From bounds on
$\mu\to e$ conversion in muon scattering off nuclei and from those on
flavour-violating hadronic tau decays the following bounds on neutrino
NSI with quarks are derived~\cite{Bergmann:2000gp}:
\begin{equation}\label{boundsQ} 
|\epsilon^q_{e\mu}| \lesssim 7\times 10^{-5}\,,\quad
|\epsilon^q_{e\tau}| \lesssim 7\times 10^{-2}\,.
\end{equation}
We take this as an order of magnitude estimate for the NSI in
propagation and detection at a neutrino factory experiment, since
there also processes with quarks are involved. For the $\mu-\tau$
channel the bounds are of order~\cite{BergmannAtm}
\begin{equation}\label{boundsMT}
|\epsilon_{\mu\tau}| \lesssim 5\times 10^{-2}
\end{equation} 
and for the NU coefficients upper bounds of order 0.1 are derived.

However, in Ref.~\cite{rossi} it has been stressed that in general no
model-independent relation exists between NSI coefficients for charged leptons
and neutrinos and only much weaker bounds of order $50\%$ are derived using
data from $e^+e^-$ colliders.  This more conservative viewpoint has been
exploited to show how FC neutrino interactions provide an excellent
description of the solar neutrino data, while consistent with the oscillation
description of the atmospheric data~\cite{Guzzo:2001mi}.

\section{The appearance rate in a neutrino factory experiment}
\label{sec:rate}

Let us consider the impact of NSI in source, propagation and detection
on the $e\to\mu$ channel\footnote{For the sake of clarity we keep the
  oscillation-inspired terminology ``$e\to\mu$ appearance channel''
  (``$\mu\to\mu$ disappearance channel'').  We actually mean the
  production of a wrong-sign (like-sign) muon in the detector,
  respectively.} at a neutrino factory.  Starting from the decay of a
$\mu^+$, we make the following ansatz for the rate at which a neutrino
produced together with the positron leads to the production of a
$\mu^-$ in the detector \cite{ota}:
\begin{equation}\label{rate}
\mathcal{R}_{e\mu} = \left| \sum_{\alpha\beta}
\mathcal{A}^S_{e\alpha} 
\mathcal{A}^P_{\alpha\beta}
\mathcal{A}^D_{\beta \mu} \right|^2 \,,
\end{equation}
and similar for the charge conjugated process.
Here we define the amplitudes describing the neutrino source and detection
process as
\begin{equation}\label{ampl}
\mathcal{A}^X_{\alpha\beta} \equiv 
\delta_{\alpha\beta}+\epsilon^X_{\alpha\beta}
\quad\mbox{for}\quad X=S,D\,,
\end{equation}
and the amplitude $\mathcal{A}^P_{\beta\gamma}$ describes the
propagation of the neutrino state from the production point to the
detector.  This amplitude is obtained from the solution of a Schr{\"o}dinger
equation with the Hamiltonian
\begin{equation}\label{ham}
H_\nu = \frac{1}{2E_\nu} \,
U \mbox{diag}\,(0,\Delta m^2_{\mathrm{sol}},\Delta m^2_{\mathrm{atm}})\,U^\dagger +
\mbox{diag}(V,0,0) + 
\sum_f V_f\,\epsilon^f \,,
\end{equation}
which takes into account neutrino oscillations and SM interactions as
well as NSI with the matter crossed by the neutrino beam.
Here $E_\nu$ is the neutrino energy and $V=\sqrt{2}G_F N_e$ is the
matter potential due to the SM charged current interaction~\cite{MSW},
where $N_e$ is the electron number density.  The last term in
Eq.~(\ref{ham}) describes the NSI with earth matter.  The sum is over
all fermions $f$ present in matter, and
$V_f\,\epsilon^f_{\alpha\beta}$ is the coherent forward scattering
amplitude of the process $\nu_\alpha + f \to \nu_\beta + f$, where
$V_f=\sqrt{2} G_F N_f$, with the number density of the fermion $f$
along the neutrino path given by $N_f$. We define an effective NSI
coefficient for the propagation by normalizing all contributions to
the down-quark potential $V_d$:
\begin{equation}\label{defeP}
\epsilon_{\alpha\beta}^P \equiv
\sum_f \frac{V_f}{V_d} \, \epsilon^f_{\alpha\beta} \,.
\end{equation}

Adopting a basis where the charged lepton mass matrix is diagonal, the
unitary matrix $U$ in Eq.~(\ref{ham}) relates the neutrino fields in
the basis where the neutrino mass matrix is diagonal to the neutrino
fields in the basis where the interaction with the SM W-boson is
diagonal~\cite{Grossman:1995wx}. We parameterize this matrix in the 
following way~\cite{theory}:
\begin{equation}
U=U_{23}U_{13}U_{12}=
\left(\begin{array}{ccc}
1&0&0\\ 0 & c_{23} & s_{23}\\ 0 & -s_{23} & c_{23} 
\end{array}\right)
\left(\begin{array}{ccc}
c_{13}&0 & s_{13} \\ 0 & 1 & 0 \\ -s_{13}&0 & c_{13} 
\end{array}\right)
\left(\begin{array}{ccc}
c_{12} & e^{i\delta}s_{12}& 0\\ -e^{-i\delta}s_{12} & c_{12}&0\\0&0&1 
\end{array}\right)\,,
\end{equation}
where $s_{ij}=\sin\theta_{ij}$ and $c_{ij}=\cos\theta_{ij}$.

In this paper we consider the following simplified scenario.  First,
we take all $\epsilon^X_{\alpha\beta}$ real and we assume that they
are the same for neutrinos and anti-neutrinos as in \cite{Val}.
Second, Eqs.(\ref{boundsLep}) and (\ref{boundsQ}) suggest that
constraints on FC interactions in the $e-\mu$ channel are about 3
orders of magnitude stronger than in the other channels.  This
motivates the approximation\footnote{Note that in
  Ref.~\cite{Gonzalez-Garcia:2001mp} much weaker sensitivities on FCI
  in the $e-\mu$ channel for the source at a neutrino factory are
  derived, of the order of $10^{-3}$. A non-vanishing
  $\epsilon_{e\mu}^S$ would imply yet an additional source for the
  confusion of oscillations and NSI.}
\begin{equation}\label{emuapprox}
\epsilon^X_{e\mu} \approx \epsilon^X_{\mu e} \approx 0
\quad\mbox{for}\quad X=S,\, P,\, D \,. 
\end{equation}
Third, we neglect the solar mass-squared difference, which implies
also that the angle $\theta_{12}$ and the phase $\delta$ disappear
\cite{theory}.  Then we are left with the following  neutrino
propagation Hamiltonian 
\begin{equation}\label{ham2}
H_\nu = 
U_{23} U_{13}\,\mbox{diag}\,(0,0,\Delta)\,U^\dagger_{13}U^\dagger_{23} +
\mbox{diag}(V,0,0) + 
V r\,\epsilon^P \,,
\end{equation}
where we have defined $\Delta\equiv\Delta m^2_{\mathrm{atm}}/2E_\nu$
and $r\equiv V_d/V=N_d/N_e$ with $r\approx 3$ in earth matter.  In the
Hamiltonian (\ref{ham2}) a sign change of $\Delta$ is equivalent to a
sign change of $V$, which interchanges the evolution of neutrinos and
anti-neutrinos.  Therefore, it is sufficient to consider only the case
$\Delta > 0$, assuming that the neutrino factory is run in both
polarities.

A detector close to the front end of a neutrino factory -- a so-called
{\sl near detector} -- can be a very powerful tool to constrain
NSI~\cite{Datta:2000ci,Bueno:2001jy,Mangano:2001mj}. Such a detector has to be
situated at a short distance (a few 100 m)
from the production region of the neutrinos, such that no oscillations
with $\Delta m^2_\mathrm{atm}$ or $\Delta m^2_\mathrm{sol}$ can develop
and matter effects are negligible. In our formalism this means that
$\mathcal{A}^P_{\alpha\beta}=\delta_{\alpha\beta}$, and the transition
rate relevant for a near detector is simply given by
\begin{equation}\label{rateND}
\mathcal{R}_{\alpha\beta}^\mathrm{ND} = \left| \sum_{\gamma}
\mathcal{A}^S_{\alpha\gamma}
\mathcal{A}^D_{\gamma\beta} \right|^2 \,.
\end{equation}
It is clear that a near detector cannot provide any model-independent
information on $\epsilon^P_{\alpha\beta}$, and only a {\sl
  combination} of $\epsilon^S_{\alpha\beta}$ and
$\epsilon^D_{\alpha\beta}$ is constrained.

Some remarks are in order. Although the general motivation for NSI is that
these accompany models of neutrino mass generation, in our following
phenomenological studies we will restrict our attention only to total lepton
number conserving NSI. While this will suffice to make our point, it will on
the other hand greatly simplify our analysis. This happens because in this
particular case it is possible to distinguish the neutrino produced together
with the electron from the one produced together with the muon, if the
detector can determine the charge of the muon. We insist, however, that in a
generic theory for NSI and neutrino masses also lepton number violating
processes~\cite{Bergmann:2000gn,Bueno:2001jy} are expected, due to the
Majorana nature of neutrinos (see Sec.~\ref{sec:models}). Such effects would
be an additional source for the confusion of NSI and oscillations.

The off-diagonal elements $\epsilon^X_{\alpha\beta}\:(X=S,P,D)$ with
$\alpha\neq\beta$ describe FC, whereas the diagonal elements with $\alpha =
\beta$ lead to NU. In Eqs.~(\ref{rate}) and (\ref{rateND}) we consider only
processes with the same final states (in source and detector) as in the SM
case. If additional processes are present, with different final states, the
corresponding amplitudes have to be added incoherently to the rate~\cite{ota}.

We want to stress that our numerical results are not restricted only
to the NSI resulting from the four-fermion operators discussed in
Sec.~\ref{sec:framework}. The results apply to {\sl all kinds of
non-standard physics} in source, propagation and detection in a
neutrino factory experiment, which can be parametrized like in
Eqs.~(\ref{rate}), (\ref{ampl}), (\ref{ham2}).
In general the NSI parameter combinations involved in the quantum
mechanical evolution of the neutrino system are model-dependent
functions of the parameters appearing in the Lagrangian 
of a given theory, as discussed in Sec.~\ref{sec:models}.

\section{The oscillation--NSI confusion theorem}
\label{sec:confusion}

In this section we present analytical approximations for the
transition rate and we show that within our simplified scenario NSI
can lead to exactly the same signal at a neutrino factory as expected
from genuine neutrino oscillations due to $\theta_{13}$.  Taking into
account additional parameters like $\Delta m^2_{\mathrm{sol}}$ or CP
violating phases in the lepton mixing matrix or in NSI can only bring
more serious complications for the determination of
$\theta_{13}$~\cite{Freund:2001ui}.

 \begin{figure}[tb!]
  \begin{center}
    \includegraphics[width=0.8\textwidth]{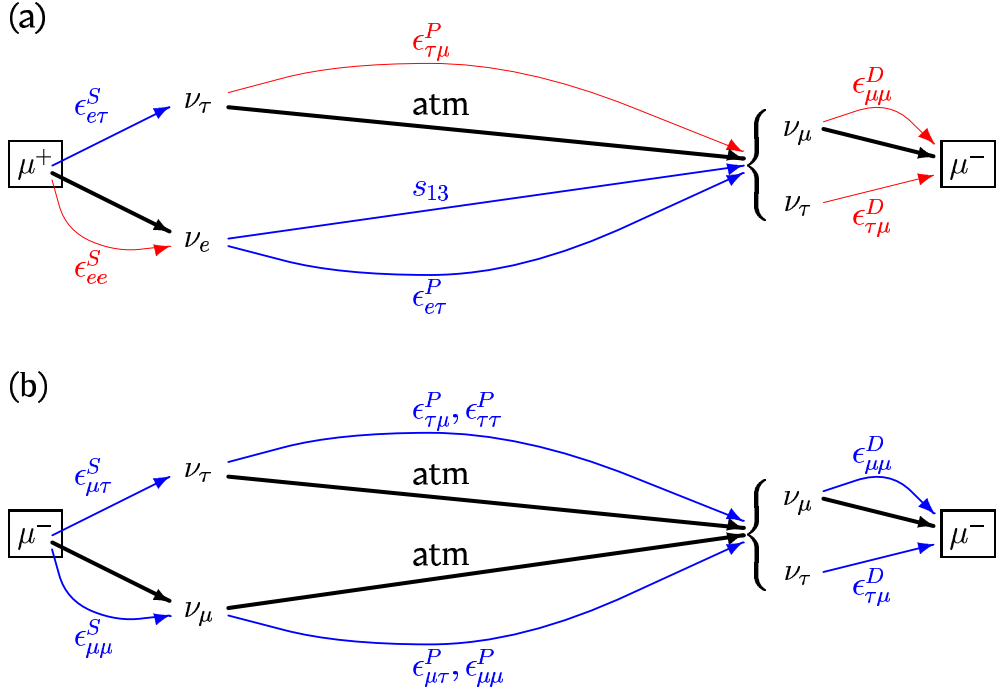}
    \caption{NSI contributions to the (a) appearance and (b) disappearance 
       channel in the approximation Eq.~(\ref{emuapprox}). 
       The thick lines indicate the dominating processes, whereas the
       thin lines show the leading contributions of the small quantities
       $s_{13}$ and $\epsilon^X_{\alpha\beta}$. The processes shown with 
       very thin lines are double suppressed in small quantities.}
    \label{fig:diagram}
  \end{center}
\end{figure}

For the understanding of the physics relevant for the numerical
results which we will present in the following sections it is useful
to derive an analytic expression for the appearance rate
Eq.~(\ref{rate}).  To this aim we assume a constant matter potential
$V$ and consider the terms containing the NSI parameters
$\epsilon^P_{\alpha\beta}$ and $s_{13}$ as a small perturbation of the
Hamiltonian and calculate eigenvalues and eigenvectors of
Eq.~(\ref{ham2}) up to first order in these small quantities.  Then
the appearance rate Eq.~(\ref{rate}) is of second order in $s_{13}$
and $\epsilon^X_{\alpha\beta}\: (X=S,P,D)$ and we make the interesting
observation that only the three parameters ($s_{13},\,
\epsilon^S_{e\tau},\, \epsilon^P_{e\tau}$) appear. 

This is a special
feature of the $e\to\mu$ channel under the approximation
(\ref{emuapprox}) and can be understood from Fig.~\ref{fig:diagram}~(a),
where we show schematically the various contributions to this channel.
The thick lines indicate the SM processes in source and detection and
the dominating $\mu\leftrightarrow\tau$ oscillations due to atmospheric
oscillation parameters. The thin lines show the leading contributions of
small quantities, which involve only the parameters $s_{13},\,
\epsilon^S_{e\tau}$ and $\epsilon^P_{e\tau}$. With very thin lines we 
show some channels which involve more than one small quantity, and
hence do not appear up to second order in the appearance rate:
we consider only the detection of muons, therefore
in leading order no effects of NSI in the detector show up because of
Eq.~(\ref{emuapprox}). Similarly no FC effects in the $\tau
\leftrightarrow \mu$ channel show up, since transitions from $e$ to
$\tau$ flavor already involve a small quantity, either $s_{13}$ or
$\epsilon^{S,P}_{e\tau}$.  We also note that no NU coefficient
$\epsilon^X_{\alpha\alpha}$ appears in leading
order~\cite{Gago:2001xg,Gonzalez-Garcia:2001mp}, explaining the lack of
sensitivity of neutrino factory experiments to NU parameters
\cite{Huber:2001zw}.  

Let us introduce the abbreviations
\begin{equation}
\epS \equiv \epsilon^S_{e\tau}\,,\quad 
\epP \equiv \epsilon^P_{e\tau}\,.
\end{equation}
Then the expression for the appearance rate is a general quadratic
form in the variables $s_{13},\,\epS$ and $\epP$:
\begin{equation}\label{prob}
\mathcal{R}_{e\mu}\approx A\,s_{13}^2 + B\,s_{13}\epP + 
C\,\epP^2 + D\,\epP \epS + E\, \epS^2 + F\, s_{13}\epS  
\end{equation}
with the coefficients
\begin{equation}\label{coeff}
\begin{array}{r@{\:=\:}l}
\rule{0mm}{6mm} A &
4\, s_{23}^2 \left(\frac{\Delta}{\Delta-V}\right)^2
      \sin^2\frac{(\Delta-V)L}{2}\,, \\
\rule{0mm}{6mm}
B & 4\, s_{23}^2 c_{23}\, r \, \frac{\Delta}{\Delta-V}
      \left[\frac{\Delta+V}{\Delta-V}\, \sin^2\frac{(\Delta-V)L}{2} +
            \sin^2\frac{VL}{2} - \sin^2\frac{\Delta L}{2} \right]\,, \\
\rule{0mm}{6mm}
C & 4\, s_{23}^2 c_{23}^2 \,r^2 \, \frac{\Delta}{\Delta-V}
      \left[\frac{V}{\Delta-V}\, \sin^2\frac{(\Delta-V)L}{2} +
            \sin^2\frac{VL}{2} - 
            \frac{V}{\Delta}\,\sin^2\frac{\Delta L}{2} \right] \,, \\
\rule{0mm}{6mm}
D & 4\, s_{23}^2 c_{23}^2 \,r \, \frac{\Delta}{\Delta-V}
      \left[\sin^2\frac{(\Delta-V)L}{2} - \sin^2\frac{VL}{2} -
      \left( 1-2\frac{V}{\Delta} \right) \sin^2\frac{\Delta L}{2} \right] \,,\\
\rule{0mm}{6mm}
E & 4\, s_{23}^2 c_{23}^2 \,\sin^2\frac{\Delta L}{2} \,,\\
\rule{0mm}{6mm}
F & 4\, s_{23}^2 c_{23} \, \frac{\Delta}{\Delta-V}
      \left[\sin^2\frac{(\Delta-V)L}{2} - \sin^2\frac{VL}{2} +
      \sin^2\frac{\Delta L}{2} \right] \,,
\end{array}
\end{equation}
where $L$ is the distance between neutrino source and detector.  The
appearance rate for anti-neutrinos $\mathcal{R}_{\bar e \bar\mu}$ is
obtained by replacing $V\to -V$ in Eq.~(\ref{coeff}).  These analytic
expressions are in agreement with numerical calculations within a few
\% in the relevant parameter range.  In general all coefficients
($A,\ldots,F$) are of the same order of magnitude and depend on neutrino
energy (via $\Delta$), on the baseline and on the sign of $V$
(neutrinos or anti-neutrinos) in a nontrivial way.

Performing a similar consideration for the $\mu$ disappearance channel
one finds that the rate $\mathcal{R}_{\mu\mu}$, defined in a way
similar to Eq.~(\ref{rate}), contains terms of all powers of the small
quantities. This is illustrated in Fig.~\ref{fig:diagram}~(b).
The zeroth order contribution corresponds to $\mu\leftrightarrow\tau$
oscillations with atmospheric neutrino oscillation parameters:
in contrast to Fig.~\ref{fig:diagram}~(a), in the case of
Fig.~\ref{fig:diagram}~(b)
there is a channel involving only thick lines.
However, we find that up to second order in small quantities only the
parameters ($\epsilon^S_{\mu\tau}, \epsilon^S_{\mu\mu},
\epsilon^P_{\mu\tau}, \epsilon^P_{\mu\mu},
\epsilon^P_{\tau\tau},\epsilon^D_{\tau\mu}, \epsilon^D_{\mu\mu}$)
appear. The important observation is that none of the three parameters
($s_{13},\epsilon^S_{e\tau},\epsilon^P_{e\tau}$) relevant for the
$e\to\mu$ channel appears. Therefore, no additional information on
these parameters can be obtained by considering the disappearance
channel. An analysis of this channel including all the parameters
listed above goes beyond the scope of this paper and we will not
consider it any further here.

Let us compare the transition rate for pure oscillations ($\epS=\epP=0$)
\begin{equation}
\mathcal{R}_{e\mu}^{\mathrm{osc}}(s_{13}) =  A\,s_{13}^2
\end{equation}
with the transition rate without oscillations ($s_{13}=0$) but
non-zero NSI coefficients
\begin{equation}
\mathcal{R}_{e\mu}^{\mathrm{NSI}}(\epS,\epP) =
C\,\epP^2 + D\,\epP \epS + E\, \epS^2 \,.
\end{equation}
With the expressions for $A,C,D,E$ 
given in Eq.~(\ref{coeff}) it is easy to check
that if the relation 
\begin{equation}\label{epsRel}
\epS = r\,\epP
\end{equation}
holds, oscillations are indistinguishable from NSI. More precisely, we obtain
\begin{equation}
\mathcal{R}_{e\mu}^{\mathrm{NSI}}\left( r\epP,\epP \right) =
\mathcal{R}_{e\mu}^{\mathrm{osc}}\left( s_{13} \right)
\end{equation}
with
\begin{equation}\label{s13Rel}
s_{13}^2 =  r^2 \epP^2 \,\frac{ 1 + \cos 2\theta_{23} }{2} \,.
\end{equation}
This means that for each value of $s_{13}$ there is a pair of NSI
parameters ($\epS,\epP$) determined by Eqs.~(\ref{epsRel}) and
(\ref{s13Rel}) which in our approximation leads to {\sl exactly the
  same} signal as oscillations due to $s_{13}$. This includes both
energy and baseline dependence, for both neutrinos and anti-neutrinos.
We call this the ``oscillation--NSI confusion theorem''.

Of course, relation (\ref{epsRel}) represents a fine-tuning of the
parameters $\epP$ and $\epS$. However, as long as this relation cannot
be excluded one has to consider this possibility. Moreover, in a
realistic experiment with finite errors and statistical uncertainties
there will be a {\sl region} around the point in the ($\epS,\epP$)
plane corresponding to Eqs.~(\ref{epsRel}) and (\ref{s13Rel}) in which
oscillations cannot be distinguished from NSI.

\section{The simulation of a neutrino factory experiment}
\label{sec:simulation}

In our numerical calculations we assume a neutrino factory with an
energy of 50 GeV for the stored muons and $2\times 10^{20}$ useful muon
decays of each polarity per year for a period of 5 years.  We consider
a magnetized iron calorimeter with a mass of $40\,\mathrm{kt}$.  The
neutrino detection threshold is set to 4 GeV, the energy resolution of the
detector is approximated by a Gaussian resolution function with $\Delta
E_\nu/E_\nu=10\%$ and we use 20 bins in neutrino energy.  We do not include
any backgrounds, efficiencies and errors in the particle
identification.  The amplitude $\mathcal{A}^P_{\alpha\beta}$ describing the
neutrino propagation is obtained by numerically solving the neutrino
evolution equation with the Hamiltonian (\ref{ham2}), using the
average matter density along each baseline.  In
Ref.~\cite{Freund:1999vc} it was shown that this is an excellent
approximation as long as the baseline is shorter than approximately
$10\,000\,\mathrm{km}$, {\sl i.e.}~as long as it does not cross the
core.  Then the transition rate Eq.~(\ref{rate}) is folded with
neutrino flux, cross section and energy resolution function to in
order obtain the expected event rates in the detector.  For further
details see Refs.~\cite{Freund:2001ui,FHL}.

Our ``observables'' are the event rates for the appearance channel
$n^i_\nu \:(n^i_{\bar\nu})$ for neutrinos (anti-neutrinos) in each
energy bin $i$. We fix the atmospheric oscillation
parameters\footnote{They will be determined at the neutrino factory
  with high accuracy from the disappearance channel.}  at their best
fit values given in \cite{Gonzalez-Garcia:2001sq,Val} $\Delta
m^2_\mathrm{atm} = 3 \times 10^{-3}\,\mathrm{eV}^2$ and $\sin^22\theta_{23}=1$.
Hence, at a given baseline, the event rates depend on the three
parameters $\si$, $\epS$ and $\epP$, where we have introduced the
abbreviation $\si\equiv\sin^22\theta_{13}$.

In order to evaluate the impact of $\epS$ and $\epP$ on the capability
of a neutrino factory to measure $\si$ we proceed as follows.  To test
a given point $(\si^0,\,\epS^0,\epP^0)$ in the parameter space we
calculate the event rates $(n_x^i)^0\equiv n_x^i(\si^0,\,\epS^0,\,\epP^0)$
with $x=\nu,\,\bar\nu$ and construct a $\chi^2$ appropriate for a Poisson
distribution\footnote{We are dealing with a counting experiment with
  eventually very low counts.}:
\begin{equation}\label{chi2}
\begin{aligned}
\chi^2(\si,\epS,\epP; &\, \si^0,\epS^0,\epP^0) = \\
&2 \sum_{x=\nu,\bar\nu} \sum_i
\left[n_x^i(\si,\epS,\epP) - (n_x^i)^0 + 
(n_x^i)^0 \, \ln\frac{(n_x^i)^0}{n_x^i(\si,\epS,\epP)}\right]\,.
\end{aligned}
\end{equation}
Thus we obtain an allowed region in the three dimensional space
$\mathcal{P}$ spanned by $(\si,\epS,\epP)$ in the usual way. Note that
the minimum of the $\chi^2$ defined in Eq.~(\ref{chi2}) is zero and
occurs at $(\si,\epS,\epP)=(\si^0,\epS^0,\epP^0)$. Therefore, the
allowed region at the CL $\alpha$ is given by the set of all points in
$\mathcal{P}$ which fulfill
\begin{equation}
\chi^2(\si,\epS,\epP;\si^0,\epS^0,\epP^0)\leq \Delta\chi^2_{\alpha} \,,
\end{equation}
where $\Delta\chi^2_\alpha$ is determined by a $\chi^2$-distribution
with 3 degrees of freedom. For simplicity we consider only starting
values with $\epS^0=\epP^0=0$, {\sl i.e.}~pure oscillations.

\section{Sensitivity limits for $\sin^22\theta_{13}$}
\label{sec:sensitivity}

 \begin{figure}[tb!]
  \begin{center}
    \includegraphics[width=0.35\textwidth,angle=-90]{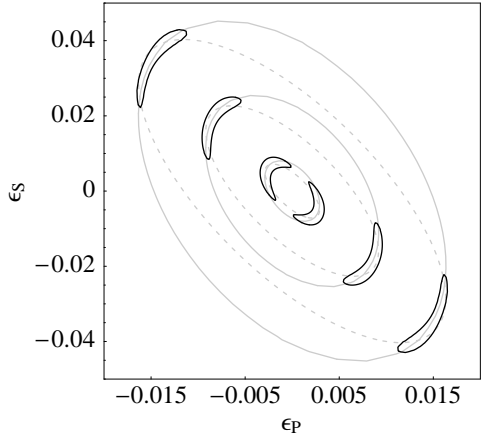}
    \caption{Allowed regions at 90\%~CL (black solid lines) in the
      $\si\equiv 0$ plane for three different starting values $\si^0 =
      (3.2\times 10^{-3},\, 10^{-3},\, 10^{-4})$ and $\epS^0=\epP^0=0$
      always.  The baseline is $3\,000\,\mathrm{km}$. The gray lines
      indicate points with the same event rates as the starting point
      (gray solid for neutrinos, gray dashed for anti-neutrinos).}
    \label{fig:th13}
  \end{center}
\end{figure}

In Fig.~\ref{fig:th13} we show the allowed regions in the $\si\equiv
0$ plane for a baseline of $3\,000\,\mathrm{km}$ and three different
starting values for $\si^0$. In gray we show the lines with the same
total event rates (solid for neutrinos, dashed for anti-neutrinos) as
in the starting point.  The general shape of these lines can
immediately be understood from Eq.~(\ref{prob}).  The small regions
delimited by the dark solid lines arise from a global fit procedure
including also the information on the spectra with the expected energy
resolution as described above. One notices that these confidence
regions follow closely the intersection of the lines of constant rates
for neutrinos and anti-neutrinos.
This means that most information is obtained from simultaneously
taking into account neutrino and anti-neutrino
rates~\cite{Huber:2001de}.  This follows from the fact that the
allowed regions extend as long as the lines of constant neutrino and
anti-neutrino rate are close to each other, {\sl i.e.~both} rates are
similar to the ones in the test point.  We conclude that it is
important to run the neutrino factory in both polarities.
On the other hand we learn that most of the information is contained
in the total rates; the spectral information is not very important:
our results are rather insensitive to variations of the number of
energy bins and of the energy resolution assumed.

 \begin{figure}[tb!]
  \begin{center}
    \includegraphics[width=.3\textwidth,angle=-90]{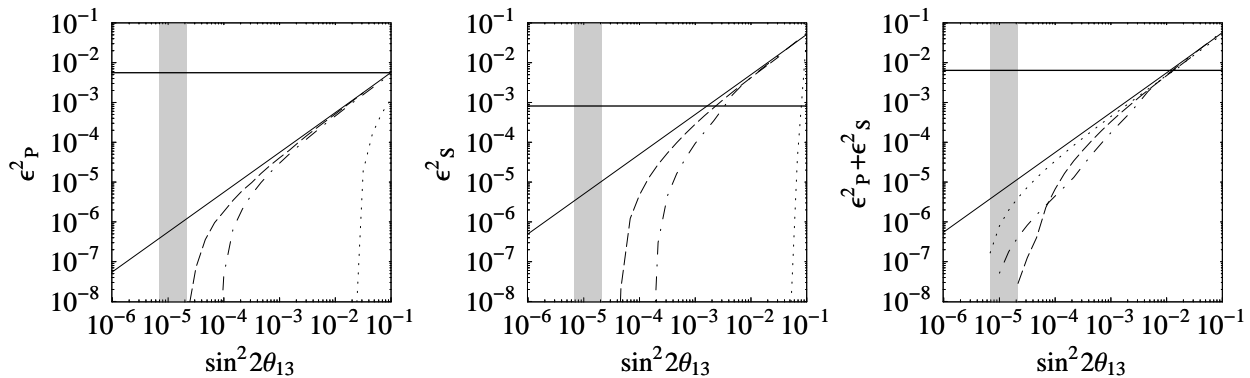}
    \caption{Sensitivity limits at 90\%~CL on $\sin^22\theta_{13}$
      attainable if a bound on $\epP^2$ (left panel), $\epS^2$ (middle
      panel) or $\epP^2+\epS^2$ (right panel) is given.  The dotted
      line is for a baseline of $700\,\mathrm{km}$, the dash-dotted
      for $3\,000\,\mathrm{km}$ and the dashed line for
      $7\,000\,\mathrm{km}$.  The horizontal black line illustrates
      the order of magnitude of current limits on the NSI parameter in
      order to ``guide-the-eye''.  The vertical gray band shows the
      range of possible sensitivities without
      NSI~\cite{Freund:2001ui}.  The diagonal solid line is the
      theoretical bound derived from our confusion theorem.}
    \label{fig:single}
  \end{center}
\end{figure}

Solutions in the $\si\equiv0$ plane of the type as shown in
Fig.~\ref{fig:th13} {\sl always} exist, irrespective of the starting
value $\si^0$. This is a consequence of the confusion theorem we have
presented in Sec.~\ref{sec:confusion}.  However, the magnitude of the
required NSI parameters $\epS$ and $\epP$ strongly depends on the size
of $\si^0$ as can be seen from Eq.~(\ref{s13Rel}) or
Fig.~\ref{fig:th13}. Thus, {\sl it is only possible to derive a limit
  on $\si$ if there is a limit on $\epS$ and/or $\epP$}; a neutrino
factory can only test a certain value of $\si$ if the values of $\epS$
and/or $\epP$, which lead to the same signal, are ruled out by some
other measurement. In Fig.~\ref{fig:single} we show the attainable
sensitivity limit on $\si$ as a function of different limits on the
NSI parameters (the present estimated NSI limits are indicated by the
solid horizontal lines). We define this sensitivity on $\si$ in the
following way.  For a dense grid of starting values $\si^0$ in the
range $10^{-6}-10^{-1}$ we calculate the 90\% CL allowed region in the
$\si\equiv0$ plane as shown in Fig.~\ref{fig:th13}.  For each value of
$\si^0$ we show the minimum value of $\epP^2$ (left panel), $\epS^2$
(middle panel) or $\epP^2+\epS^2$ (right panel) inside this allowed
region.  The neutrino factory is sensitive to this value of $\si$ only
if there is a bound on this NSI parameter (combination of parameters),
which is smaller than this minimum value.
 
Any experiment ({\sl e.g.}~like a near detector at a neutrino factory)
trying to measure $\epS$ and $\epP$ will only restrict a certain
combination of the NSI parameters in the source and the detector used
in this particular experiment (see Eq.~(\ref{rateND})). In general it
will be very difficult to translate such a result into a bound on
$\epS$, and even more difficult on $\epP$, in a model-independent way.
It is however to be expected that the constrained combination depends
on the square of the NSI parameters since the transition rate
$\mathcal{R}\propto\epsilon^2$.  Therefore we show the results for the three
simple functions $\epS^2,\, \epP^2$ and $\epS^2 + \epP^2$ as mentioned above.  
The left hand panel of Fig.~\ref{fig:single} shows the sensitivity limit
if there is a bound on $\epP^2$ and all values of $\epS$ are allowed.
It seems very difficult to obtain such a bound in a model-independent
way, because it is not possible to probe directly the NSI parameters
relevant in neutrino propagation. As recently stressed in
Ref.~\cite{rossi} in general it is not possible to use bounds on
similar processes involving charged leptons. Moreover, many different
processes may contribute to $\epP$ (see Eq.~(\ref{defeP})).  In the
middle panel we show the sensitivity limits for a bound on $\epS^2$
and all values allowed for $\epP$, which is probably the most
realistic case because it should be possible to constrain $\epS$ with
a near detector setup. In the right hand panel we display the optimal
situation, if a bound on the combination $\epS^2+\epP^2$ is available.

The diagonal solid line in Fig.~\ref{fig:single} shows the theoretical
bound implied by the oscillation--NSI confusion theorem
Eqs.~(\ref{epsRel}) and (\ref{s13Rel}). This bound corresponds to the
best possible situation, which can be achieved only for large values
of $\si$ due to large event numbers. For smaller values of $\si$ the
realistic bound gets worse because of statistical limitations due to small
event numbers.  The numerical differences between the three plots in
Fig.~\ref{fig:single} is due to the different symmetries of the used
function of $\epS$ and $\epP$ with respect to the symmetry of the
allowed regions in the $\si\equiv 0$ plane.  For small values of
$\epS^2+\epP^2$ (right panel) the bounds converge to the sensitivity
limits obtained without taking into account NSI~\cite{Freund:2001ui}.
The range of these limits for the three different baselines is shown
as the gray vertical band.

\begin{figure}[tb!]
  \begin{center}
    \includegraphics[width=.4\textwidth,angle=-90]{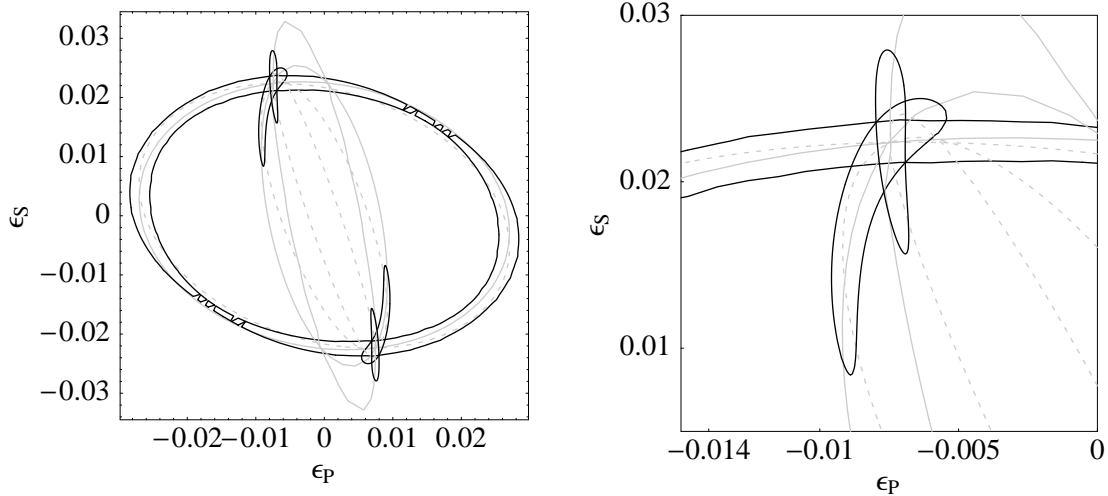}
    \caption{Allowed regions at 90\%~CL (black solid lines) in the
      $\si\equiv0$ plane for three different baselines
      ($L=1\,000\,\mathrm{km}$, $3\,000\,\mathrm{km}$,
      $5\,000\,\mathrm{km}$) and $\si^0 = 10^{-3},\:\epS^0=\epP^0=0$.
      The gray lines indicate points with the same event rates as the
      starting point (gray solid for neutrinos, gray dashed for
      anti-neutrinos). The right hand panel is a blow up of the left
      hand panel.}
    \label{fig:bases}
  \end{center}
\end{figure}

We can understand the behavior of the sensitivity limits in
Fig.~\ref{fig:single} by considering the allowed regions in the
$\si\equiv 0$ plane for different baselines, as shown in
Fig.~\ref{fig:bases}.  For small baselines the allowed region is
roughly a circle.  Therefore a bound on an individual $\epP$ or $\epS$
is useless for a sensitive determination of $\si$ (see left and middle
panel of Fig.~\ref{fig:single}).  We conclude that if only a baseline
of 700 km is available it is mandatory to establish solid bounds on
{\sl both} NSI parameters.  However, a bound on $\epP^2+\epS^2$ is most
suitable for small baselines and in this case $L=700$ km can do even
better than longer baselines (right panel of Fig.~\ref{fig:single}).
For longer baselines the allowed regions in the $\si\equiv 0$ plane become
smaller (see Fig.~\ref{fig:bases}) and hence, also a bound on an individual 
NSI parameter is useful.

With the solid horizontal lines in Fig.~\ref{fig:single} we illustrate
the order of magnitude of current bounds on NSI parameters. For $\epP$
we show the bound given in Eq.~(\ref{boundsQ}), while for the $\epS$
case we use the bound given in Eq.~(\ref{boundsLep}).  It is clearly
visible that using even these rather optimistic bounds on the relevant
NSI parameters the sensitivity of a neutrino factory is at its best
$\si \sim 10^{-3}$ -- compared to $\si\sim 10^{-5}$ - in the absence
of any NSI.  The sensitivity is deteriorated by two orders of
magnitude for all baselines. Let us stress again that we are {\sl not}
using the bounds from Eqs.~(\ref{boundsQ}) and (\ref{boundsLep}) in
our analysis because they are derived under some model-dependent
assumptions from non-neutrino processes. The horizontal lines in
Fig.~\ref{fig:single} should merely ``guide-the-eye'' in reading the
plots, they simply give a rough idea of the order of magnitude of
existing bounds.

The geometry of the currently discussed muon storage rings offers the
striking possibility to illuminate two detectors at different
baselines with neutrinos from one neutrino factory. With this in mind,
let us investigate to which extent the sensitivities for $\si$ can be
improved by combining the information of two baselines.
From Fig.~\ref{fig:bases} we find that although the shape of the
allowed regions in the $\si\equiv 0$ plane is very different for different
baselines, they all have a common intersection. Moreover, considering
the lines in the $\si\equiv 0$ plane which have the same event rates as
the starting value $\si^0$ (shown as gray solid lines for neutrinos
and gray dashed lines for anti-neutrinos) we observe that all
iso-rate lines meet in a single point.  This is again due to our
oscillation--NSI confusion theorem: Eqs.~(\ref{epsRel}) and
(\ref{s13Rel}) do not depend on the baseline.  Thus even the
combination of baselines cannot lift this degeneracy.

\begin{figure}[tb!]
  \begin{center}
    \includegraphics[width=.3\textwidth,angle=-90]{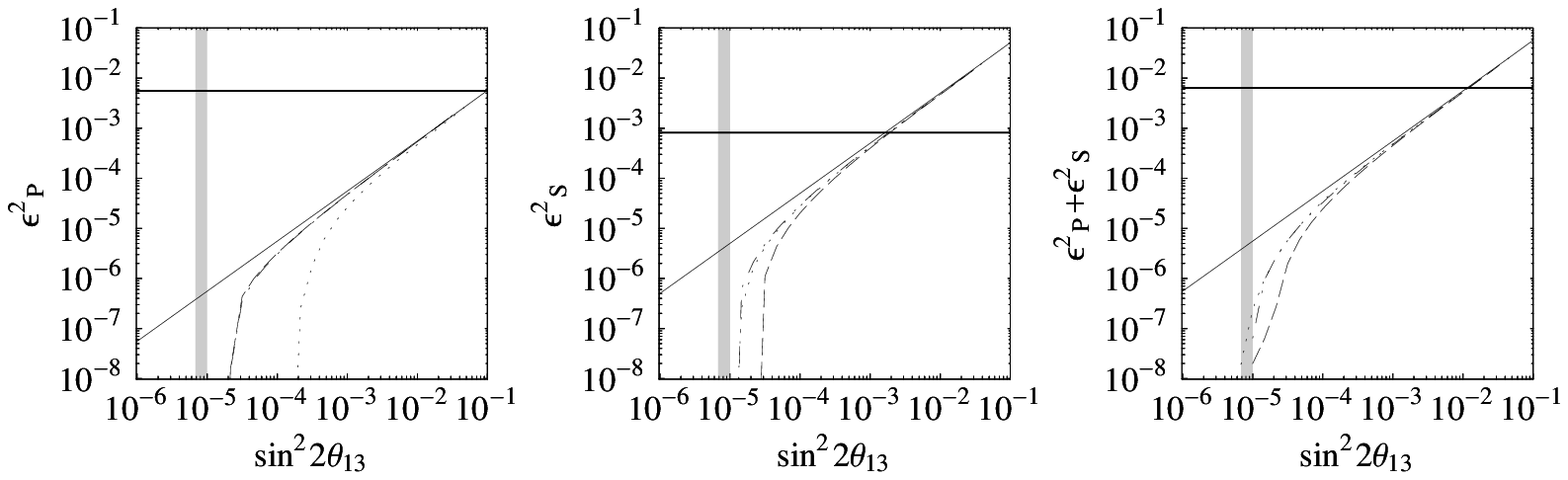}
    \caption{Sensitivity limits at 90\%~CL on $\sin^22\theta_{13}$
      attainable if a bound on $\epP^2$ (left panel), $\epS^2$ (middle
      panel) or $\epP^2+\epS^2$ (right panel) is given.  The dotted
      line is for the baseline combination
      $700\,\mathrm{km}\,\&\,3\,000\,\mathrm{km}$, the dash-dotted for
      $700\,\mathrm{km}\,\&\,7\,000\,\mathrm{km}$ and the dashed line
      for $3\,000\,\mathrm{km}\,\&\,7\,000\,\mathrm{km}$.  The
      horizontal black line illustrates the order of magnitude of
      existing limits on the NSI parameter.  The vertical gray band
      shows the range of possible sensitivities without
      NSI~\cite{Freund:2001ui}.  The diagonal solid line is the
      theoretical bound derived from the confusion theorem.}
    \label{fig:double}
  \end{center}
\end{figure}

In Fig.~\ref{fig:double} we show sensitivity limits for $\si$ obtained
as before, but now we use the sum of the $\chi^2$-functions
Eq.~(\ref{chi2}) for two different baselines. In comparison with
Fig.~\ref{fig:single} we observe that the theoretical bound can now be
achieved somewhat easier. However, in contrast to the case considered
in our previous work~\cite{Huber:2001de}, where only NSI with the
earth matter are taken into account, in our present more realistic
situation including also effects in the neutrino source even a
combination of baselines does not resolve the confusion problem. With
the current bounds on the relevant NSI parameters the sensitivity is
$\si \gtrsim 10^{-3}$, which coincides with the sensitivity obtained at a
single baseline.  Again this sensitivity is two orders of magnitude
worse than without NSI.

\section{Conclusions}
\label{sec:conclusions}

In this paper we have considered the impact of non-standard neutrino
interactions on the determination of neutrino mixing parameters at a
neutrino factory.  In particular we have focused on the so-called
``golden channels'' for the measurement of $\theta_{13}$, namely the
$\pnu{e}\to\pnu{\mu}$ channels.  We have extended our previous work
\cite{Huber:2001de} by taking into account both the effects of
neutrino oscillations as well as the effect of NSI at the neutrino
source, propagation and detection~\cite{ota}. Within a very good
approximation we have explicitly demonstrated how a certain combination
of FC interactions in source and propagation can produce exactly the
same signal as would be expected from oscillations arising due to
$\theta_{13}$.
This implies that information about $\theta_{13}$ can only be obtained if
bounds on NSI parameters are available and that all one can achieve at
a neutrino factory is a {\sl correlated oscillation--NSI study}.  In view of
the current estimates of the bounds on FC interactions, this leads to
a drastic loss in sensitivity in $\theta_{13}$, at least two orders of
magnitude.

In order to improve the situation it is mandatory to obtain better
bounds on $\epS$ and $\epP$ at the $\epsilon\simeq 10^{-4}-10^{-3}$
level, which is several orders of magnitude more stringent than
current limits.  This unexpected complication should be taken into
account in the design of a neutrino factory.
On the other hand, a neutrino factory may also offer the possibility
to obtain these very stringent limits on the NSI parameters.  Using a
small near detector ($L\approx 100\,\mathrm{m}$) with very good
particle identification for taus it would be possible to restrict the
$\mathcal{R}_{e\tau}$ down to $10^{-8}-10^{-6}$. Since
$\mathcal{R}_{e\tau}\propto\epsilon^2$ this translates into a bound
for $\epsilon\simeq 10^{-4}-10^{-3}$~\cite{Mangano:2001mj}.  Thus the
near site physics program of a neutrino factory is a necessary and
very important part of the long baseline program.

\section*{Acknowledgments}

This work was supported by Spanish DGICYT under grant PB98-0693, by
the European Commission grants HPRN-CT-2000-00148 and HPMT-2000-00124,
and by the European Science Foundation \emph{Neutrino Astrophysics
  Network} grant N.~86.

\end{document}